**In Defense of an Accelerating Universe: Model Insensitivity of the Hubble Diagram**

H. I. Ringermacher* and L. R. Mead
Dept. of Physics and Astronomy, U. of Southern Mississippi, Hattiesburg, MS 39406, USA

**ABSTRACT**

A recently published paper by Nielsen, Guffanti & Sarkar argues that the evidence for cosmic acceleration is marginal and that a coasting universe - the "Milne Universe" - fits the same supernovae data in a Hubble diagram nearly as well. The Milne Universe has negative spatial curvature. Nevertheless, we confirm that the Milne model fits the data just as well as ΛCDM. We show that this unexpected result points to a weakness in the Hubble diagram rather than to a flaw in ΛCDM. It seems the Hubble diagram is insensitive to spatial curvature. Here we present the data and both models in a scale factor vs. cosmological time plot. This plot is exquisitely sensitive to spatial curvature because one of three unique transformations, for each of $k = 0, +1, -1$ is applied to transform from the Hubble diagram. Although the Milne negative curvature did not matter much in the Hubble diagram, it matters critically in the scale factor plot. Given that space is flat as measured by precise CMB observations, we find that when the SNe data, the ΛCDM model and the Milne model are plotted as scale factor vs. cosmological time the two resulting curves separate at 2 sigma above the noise- ten times their separation above the noise in the Hubble diagram. The transformed data fit to the ΛCDM model confirms, at a 95% confidence level, that the universe is accelerating and the Milne coasting universe is ruled out.

*Communications to: ringerha@gmail.com

**1. INTRODUCTION**

Nielsen, Guffanti & Sarkar (2016), referred to as NGS, present a detailed argument that the evidence for cosmic acceleration is marginal and that a coasting model - namely that of the "Milne Universe", a universe expanding at constant speed - fits the same SNe Ia data nearly as well. They plot the usual Hubble diagram presenting distance modulus vs. redshift. The Milne Universe is an expanding space-time with scale factor $a(t) = t$, where *t* is cosmological time. It has spatial curvature $k = -1$, and equation of state $p = -\rho/3$, where *p* is pressure and $\rho$ is energy density. When the Friedmann equations for the Milne metric are converted to distance modulus vs. redshift, the resulting curve nearly overlaps the standard ΛCDM curve in the Hubble diagram. In an earlier paper (Ringermacher & Mead, 2014) we showed how to transform Hubble diagram data (distance modulus vs. redshift ) to scale factor vs. cosmological time in a "model-independent" way for a flat space. When the ΛCDM model and Milne model are plotted as scale factor vs. linear cosmological time, the two resulting curves separate significantly. In this plot, the Milne model is a diagonal straight line, while ΛCDM traces the traditional acceleration curve with an inflection point at the "transition time" of the universe. The central question is where does the data land under a transformation from modulus/redshift to scale-factor/time? We shall show that the transformation is, in effect, discontinuous and depends on a prior knowledge of the spatial curvature. NGS have demonstrated that the Hubble diagram is not particularly sensitive to spatial curvature. They correctly show that the Milne hyperbolic space model fits nearly as well to the same SNe Ia data as the ΛCDM flat space model. The transformation of the data set must be done in the same way as the model curves, that is, taking into account the spatial curvature. We shall show that if space is flat the transformed data is an excellent fit to ΛCDM and if space is hyperbolic the data is an excellent fit to the Milne model. In both cases the models and fitted data separate well above the noise. The central point that determines which fit is the correct one is the knowledge of the spatial curvature. Independent measurements, in particular the CMB uniformity (de Bernardis, et al., 2000 ) and angular power spectrum (Spergel et al., 2003) are consistent with a flat space. Thus, spatial curvature, though not relevant in the Hubble diagram, is critical in the scale factor plot. A flat space therefore supports the ΛCDM accelerating universe and rules out the Milne Universe. The following sections will develop these assertions.

## 2. THE MILNE UNIVERSE

A "coasting universe" was first described by Milne (E.A. Milne, 1932). NGS state that "the Milne model refers to an equation of state $p = -\rho/3$ and should not be taken to mean an empty universe". They suggest viscosity effects in clustering can result in approximately constant velocity expansion. Nevertheless, the Milne Universe is a metric space satisfying the Einstein equations while viscosity effects require ancillary physics that is not discussed in NGS. A Milne Universe is an empty expanding universe moving at constant speed defined by its scale factor, $a(t) = t$, where t is cosmological time. Its 3-space has negative curvature, $k = -1$, and its equation of state is $p = -\rho/3$. In fact, a generalization of the Milne Universe (Ringermacher & Mead, 2005) can be written as :

$$ds^2 = dt^2 - Dt^2 \frac{dr^2}{(1 - k r^2)}, \qquad (1)$$

where we have chosen line-of-sight coordinates. We note the "*D*-factor" leading the spatial component. Setting $D = 1$ results in the usual "empty" Milne Universe – devoid of matter. However, setting $D = 1 + \beta$, where $\beta$ is a positive constant, results in a positive energy density varying with cosmological time;

$$\rho = \frac{3\beta}{(1+\beta)t^2}. \qquad (2)$$

This leaves all the Milne properties but now a non-empty space. The space still expands at constant speed $a(t) = t$. If one insists on a *metric explanation* of the expansion of the universe at constant speed, this is the best you can do. If a space expands at "nearly constant speed" as is suggested in NGS as an alternative to the accelerating universe, then it is either decelerating or accelerating and it cannot be called a Milne Universe. For the purposes of the remainder of the paper, we write the line-of-sight FRW metric in general form for the three spatial curvatures:

$$ds^2 = dt^2 - a(t)^2 d\chi^2 \begin{cases} k = 0 & \chi = r \\ k = -1 & \chi = \sinh^{-1} r \\ k = +1 & \chi = \sin^{-1} r \end{cases} \qquad (3)$$

The Milne Universe is the case $k = -1$ and $a(t) = t$. Selection of spatial curvature makes for discontinuous transformations between cosmological model parameters. It cannot be overemphasized that a Milne space has negative spatial curvature, $k = -1$, and that this will have discontinuous consequences in transformations between cosmological parameters.

## 3. TRANSFORMATION OF THE HUBBLE DIAGRAM TO A SCALE FACTOR PLOT

Plotting distance modulus against redshift is today ubiquitous. It is true that only about 70 SNe Ia data points were used altogether between Perlmutter (Perlmutter, et al., 1999) and Riess (Riess, et al., 1998) when their discovery of an accelerating universe was announced in 1998. Since then, more SNe Ia data has accumulated and confirms the early result.

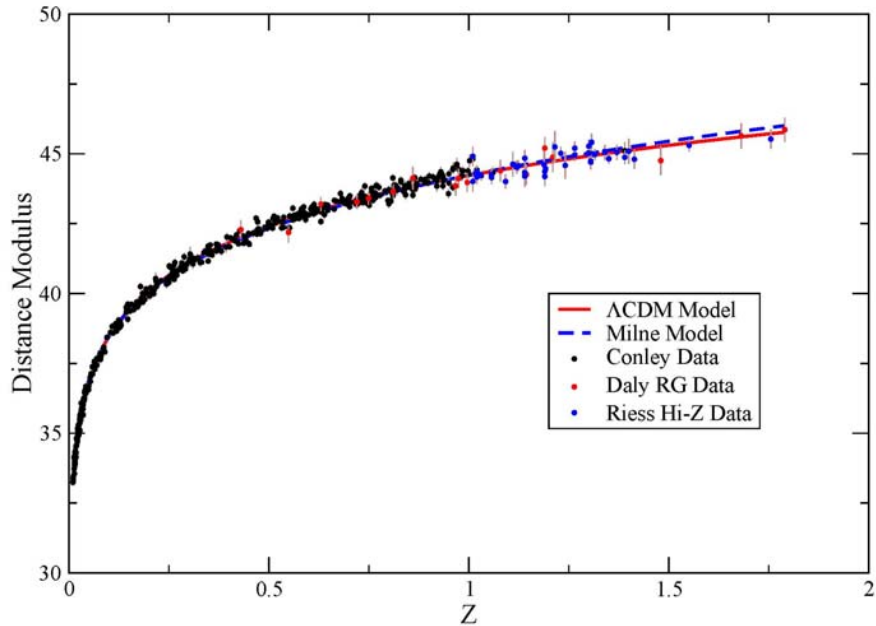

**Fig.1**. Distance Modulus vs. z. Solid curve is ΛCDM. Dashed curve is Milne model.

In Figure 1 we reproduce the Hubble diagram, distance modulus vs. redshift, for SNe Ia data (Conley et al., 2011; Riess et al., 2004) together with radio galaxy data of Daly (Guerra, Daly & Wan, 2000; Daly & Djorgovski, 2004) from our earlier paper (Ringermacher & Mead, 2014). Error bars have been added here as a reference for error propagation to the scale-factor plot. The RMS errors are comprised of quoted errors for all three data sets and include modulus error, intrinsic error and light-curve correction errors where applicable. The ΛCDM model best fit for WMAP parameters ($\Omega_\Lambda = 0.728, \Omega_m = 0.272$) is shown as the solid line. The best-fit Hubble constant is $h = 0.66$. The curve shifts within the width of a dot for Planck parameters. The dashed curve is the Milne model following the precise calculation outlined in NGS, equations (2). For the empty Milne Universe, we set $\Omega_k = 1$, $\Omega_\Lambda = \Omega_m = 0$ and $h = 0.66$. The Milne model for empty space is, in fact, an excellent fit, comparable to ΛCDM, as NGS state, although they vary multiple data parameters for their best fit. This is not a best fit here because there are no adjustable parameters in the model itself. Their examples do not fit much better than our extreme-case empty-space curve. To be more precise, the RMS modulus deviation ($\Delta_{modulus} = 0.042$) between the two model curves in Figure 1 is approximately 0.24 of the standard deviation (statistical spread) of the data ($\sigma_{modulus} = 0.175$). These statistics were obtained with no adjustment of light-curve parameters for the Milne model and a least-squares fit for the ΛCDM model. Thus the Milne coasting model is essentially as good a fit with no parameter adjustment.

To make certain this is the case we also performed a best fit for the Milne curve by optimizing the same light-curve parameters as NGS (NGS Table 1, empty universe case), in particular the color and stretch parameters. We could only do this for the Conley portion (Conley et al., 2011) of our data. This comprises about 2/3 of the same data used by NGS. We adopted the NGS Hubble constant choice $h = 0.70$, and found a best fit for absolute magnitude $M_0 = -19.00$ (NGS: -19.01), stretch parameter $\alpha = 0.98$ ( NGS: 0.13 ) and color parameter $\beta = 3.18$ ( NGS: 3.05 ). Adjustment parameters were not expected to match those of NGS for the different data sets. This best fit brought the two curves closer together to within $0.18\sigma_{modulus}$. If we define "signal" as the mean deviation between the 2 model curves, then the SNR discriminating between the two models in the Hubble diagram is at best 0.18. The Milne model is the most extreme case of NGS, Table 1, yet the Hubble diagram cannot distinguish it from ΛCDM.

In previous work (Ringermacher & Mead, 2014) we demonstrated a model-independent method of converting a Hubble diagram into a plot of scale factor vs. cosmological time. This approach did not require choosing a particular model of the universe, e.g. ΛCDM, and therefore did not require a priori knowledge of density parameters. In fact $\Omega_\Lambda$ and $\Omega_m$ are adjusted for a best fit. The transformation, however, does require a choice for the spatial curvature – in this case, a flat space for ΛCDM and a hyperbolic space for Milne.

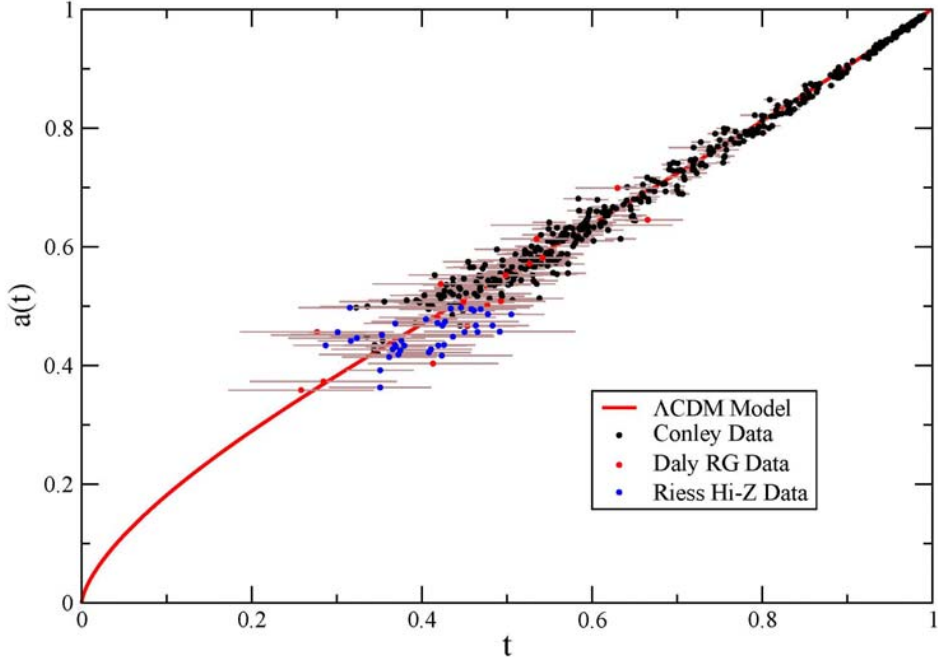

**Fig.2**. Scale factor vs. cosmological time for flat-transformed data of Figure 1. Curve is best fit ΛCDM model for $k = 0$, $\Omega_\Lambda = 0.735$, $\Omega_m = 0.265$ and $h = 0.66$.

In Figure 2 we show model-independent scale factor vs. cosmological time calculated[2] for the entire data set of Figure 1 with the best fit ΛCDM parameters, $\Omega_\Lambda = 0.735$ and $\Omega_m = 0.265$, essentially identical to the best-fit parameters of Figure 1. A flat-transform ($k = 0$) was performed on the data. Error bars for all of the data are propagated from the Hubble diagram as well and show as time-errors. The "R-square goodness-of-fit" for the ΛCDM model with the data in Figure 2 is 98.5%. This model shows an accelerating universe for SNe data as well as the Daly, et al. radio galaxy data to the highest redshift ($z = 1.79$). Scale factor is normalized to 1 at the present time, $t = 1$. The Milne model is not shown on this plot since we have used a flat-transform on the full data. We can only use the Conley data to make comparisons with NGS which will be presented in figures 3-5. Figure 2 was obtained by following the transformation outlined in the model-independent plotting method (Ringermacher & Mead, 2014) for a flat space geometry assumption. The transformed data will not fit the Milne model in this scale factor plot simply because the transformation is that for a flat space – specifically, the coordinate distance, $r$, is Euclidean. On the other hand, if we assume the space is hyperbolic, then $r$ is replaced by $\sinh r$ in our approach and the data shifts over to the Milne model best fit shown in Figure 3. Indeed, it is this very feature that is responsible for the insensitivity of the Hubble diagram to curvature – namely that in calculating the luminosity distance we find that the normalized coordinate distance $r \cong \sinh r$ for virtually all of the data. Data and model curves for Figures 1-2 can be downloaded at www.ringermacher.com from "downloadable data" and are coded with "2017SR". Only the Conley data is used in Figure 3 since the same data is used in NGS. In Figure 3 one can see that the data shift is in the time direction. The mean time-separation of the 2 model curves in Figure 3 (between cosmological times 0.4 and 1.0) is $\Delta_{time} = 0.033$. This is approximately $2\sigma_{ScaleF}$, where $\sigma_{ScaleF} = 0.017$ is the average standard time deviation of the Milne data/model and the ΛCDM data/model. By the previous definition, the SNR for the scale factor plot model separation is 2. *Thus the SNR of the scale factor plot is approximately 10 times the SNR of the Hubble diagram in separating the two models.*

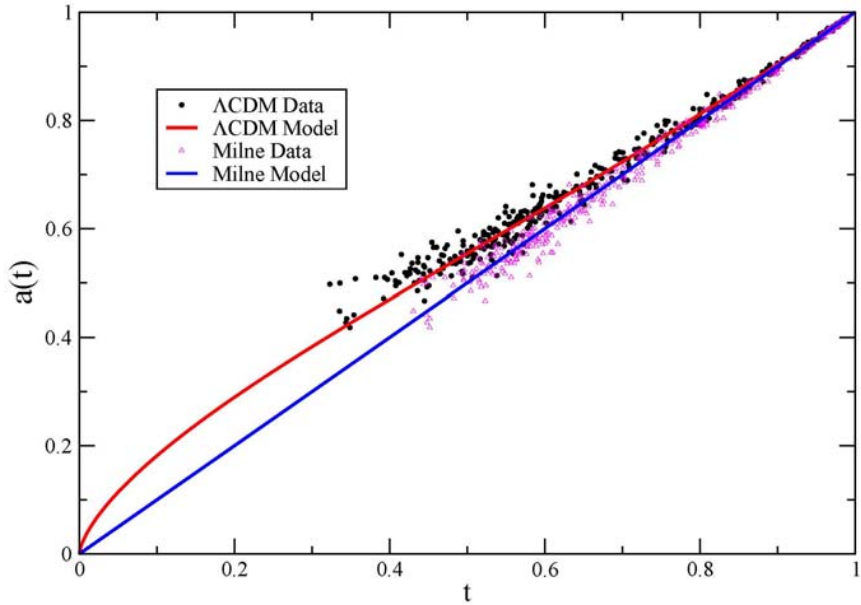

**Fig.3**. Scale factor vs. cosmological time showing data shift from the best fit ΛCDM to the best fit Milne model upon a hyperbolic transformation from the Hubble diagram.

Error bars are shown separately in Figures 4 and 5 so as to clearly display the separation of the two models. The data error is completely dominated by the time error propagated from the Hubble diagram error. That is due to the fact that redshift errors are extremely small, typically $\Delta z = 0.001$ or less. The greatest error for $z > 1$ is $\Delta z = 0.02$. This translates into a scale factor error $\Delta a \cong 0.005$. The error in a(t) is less than or equal to the dot size. The time error bars shown are the experimental measurement RMS error propagated directly from the effective distance modulus error sources arising from quoted errors and parameters (Conley et al., 2011) in apparent magnitude, m, color parameter, c, and stretch parameter, s, according to:

$$\mu_{eff} = m - M_0 + \alpha(s-1) - \beta c \qquad (4)$$

$M_0$, $\alpha$ and $\beta$ are the values quoted earlier for a best fit of the Milne a(t) model.

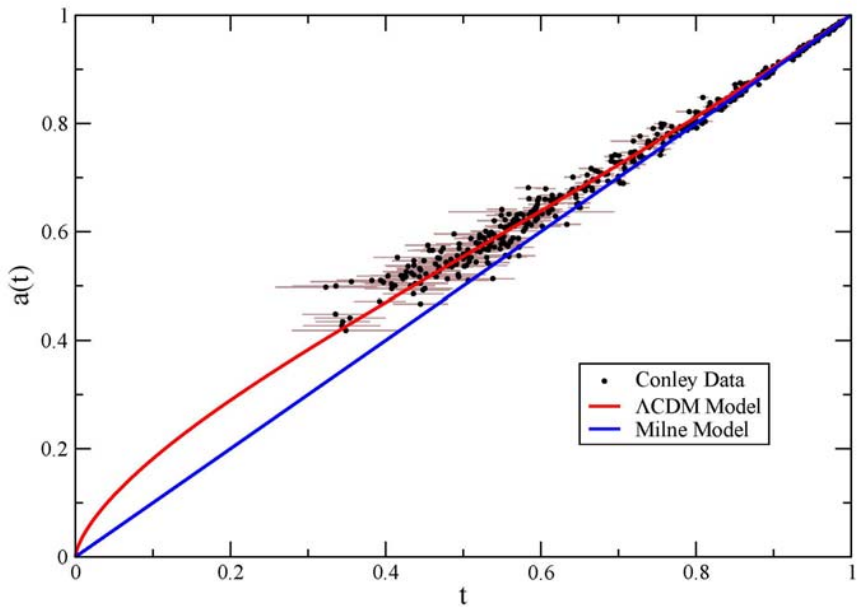

**Fig.4**. Scale factor vs. cosmological time showing best fit ΛCDM model. Space curvature is assumed flat. The Milne model (straight line) is shown for comparison.

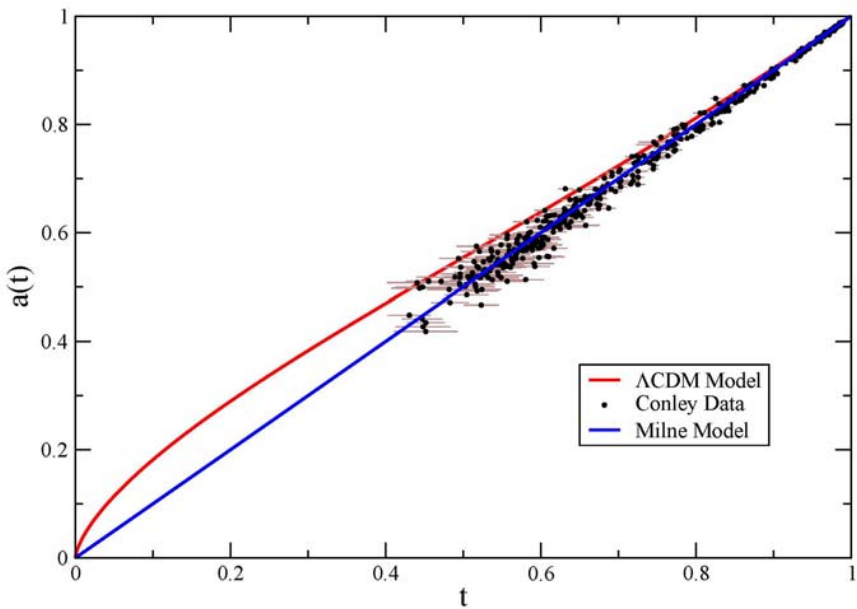

**Fig.5**. Scale factor vs. cosmological time showing best fit Milne model. Space curvature is assumed negative. The ΛCDM model (curved line) is shown for comparison.

The "R-square goodness-of-fit" for the Conley data with Milne model in Figure 5 is 99.1%.

## 4. CONCLUSIONS

At first sight, it appears that, on a Hubble diagram of distance modulus vs. redshift, an accelerating universe is indistinguishable from a coasting universe. NGS has taken this to mean that the evidence for an accelerating universe is marginal. However, we have shown that the Hubble diagram is actually at fault: it is fundamentally insensitive to spatial curvature. By transforming the SNe data to *model-independent* scale factor vs. cosmological time using methods described in previous work (Ringermacher & Mead, 2014), we have shown that the ΛCDM model and Milne model clearly separate with a ten-fold SNR improvement based on spatial curvature. The ΛCDM scale-factor model best fit is for WMAP parameters $\Omega_\Lambda = 0.728, \Omega_m = 0.272$ but Planck Hubble parameter $h = 0.66$. Transforming to a scale-factor/time plot requires a separate transformation for each of the three spatial curvatures replacing the flat space Euclidean *r* by sinh *r* for negative curvature and by sin *r* for positive curvature. Thus a scale factor plot separates models for spaces of a given curvature. Given that independent CMB experimental evidence favors flat space, a Euclidean transformation of the SNe data to scale factor vs. cosmological time supports a best fit to the ΛCDM model. This makes it clear that the universe is accelerating at a 95% confidence level and that a hyperbolic space Milne model describing a coasting universe is ruled out.

## ACKNOWLEDGEMENTS

We wish to thank Douglas Seiler for making us aware of the Nielsen, Guffanti and Sarkar paper.